\documentclass[prd,nofootinbib,showpacs,preprint]{revtex4}
\usepackage{graphicx}
\usepackage{epsfig}
\usepackage{amsmath}
\usepackage{amsfonts}
\usepackage{amssymb}
\usepackage{url}
\usepackage{hyperref}
\usepackage{subfigure}
\usepackage{cancel}
\usepackage{eso-pic}
\usepackage[abs]{overpic}
\usepackage{xcolor,varwidth}
\newcommand{\beqa}{\begin{eqnarray}}
\newcommand{\eeqa}{\end{eqnarray}}
\newcommand{\beq}{\begin{equation}}
\newcommand{\eeq}{\end{equation}}

\newcommand{\nn}{\nonumber}
\newcommand{\bmt}{\begin{pmatrix}}
\newcommand{\emt}{\end{pmatrix}}
\usepackage[toc,page]{appendix}
\usepackage{comment}
\newcommand{\be}{\begin{equation}}
\newcommand{\ee}{\end{equation}}
\newcommand{\bea}{\begin{eqnarray}}
\newcommand{\eea}{\end{eqnarray}}

\usepackage{xspace}

\begin{document}
\title{Event Reconstruction in the NOvA Experiment}
\author{Biswaranjan Behera$^1$  }
\email{bbehera@fnal.gov}
\author{ Gavin Davies$^2$ }
\author{ Fernanda Psihas$^2$ }
\affiliation{\,$^1$Department of Physics, IIT Hyderabad,
              Telangana - 502285, India \\
$^1$Fermilab, USA\\}

\affiliation{$^2$\,Indiana University, Bloomington\\
On behalf of the NOvA Collaboration\\ \\
Talk presented at the APS Division of Particles and Fields Meeting (DPF 2017), July 31-August 4, 2017, Fermilab. C170731}
              

\begin{abstract}
The NOvA experiment observes oscillations in two channels (electron-neutrino appearance and muon-neutrino disappearance) using a predominantly muon-neutrino NuMI beam. The Near Detector records multiple overlapping neutrino interactions in each event and the Far Detector has a large background of cosmic rays due to being located on the surface. The oscillation analyses rely on the accurate reconstruction of neutrino interactions in order to precisely measure the neutrino energy and identify the neutrino flavor and interaction mode. Similarly, measurements of neutrino cross sections using the Near Detector require accurate identification of the particle content of each interaction. A series of pattern recognition techniques have been developed to split event records into individual spatially and temporally separated interactions, to estimate the interaction vertex, and to isolate and classify individual particles within the event. This combination of methods to achieve full event reconstruction in the NOvA detectors has discussed.
\end{abstract}
\maketitle
\section{Introduction}
The NOvA (NuMI Off-axis $\nu_{e}$  Appearance) far detector (FD) is 810 km from the NuMI production target and positioned 14 mrad off-axis from the NuMI beam, resulting in a narrow-band neutrino flux peaked around 2 GeV \cite{tdr}.  The NOvA near detector (ND) is located approximately 1 km from the NuMI production target, off-axis such that the peak of the neutrino flux matches that of the far detector. Both detectors are functionally identical, segmented, tracking calorimeters. The basic unit of the NOvA detectors is a long highly reflective white polyvinyl chloride (PVC) cell of cross sectional size 3.9cm by 6.6cm filled with liquid-scintillator. The detectors are designed to provide sufficient sampling of hadronic and electromagnetic showers to allow efficient separation of the charged current (CC) interaction signals from the neutral current (NC) interaction backgrounds. There are several physics goals in the NOvA experiment. These are observation of the oscillation of muon neutrinos to electron neutrinos, neutrino mass ordering and CP violation in neutrinos. In addition to that, there are searches for sterile neutrinos, supernova, and neutrino cross section measurements in the ND.

\begin{figure}[!htbp]
    \centering
       \includegraphics[width=0.9\textwidth]{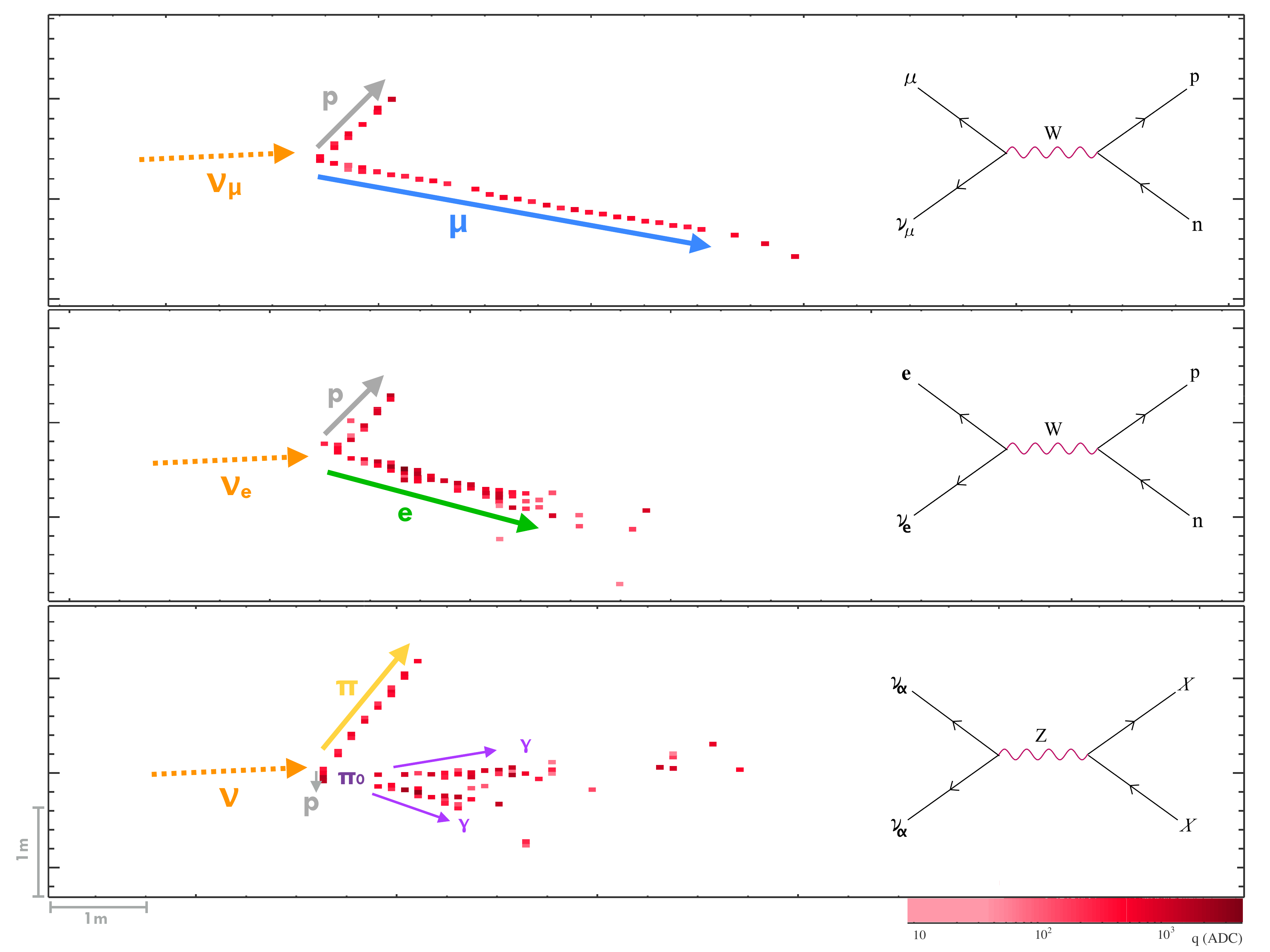}
     
    \caption{ $\nu_\mu$ CC neutrino interaction with long muon (top).  $\nu_{e}$ CC neutrino interaction with electron shower (middle),  Neutral current (bottom).}
    \label{fig:evd}
  \end{figure}

The analyses goals of the NOvA experiment require detailed reconstruction of the neutrino interactions. The signature of events (individual fundamental interactions) looks different for different interactions. The $\nu_{e}$ charged current  interaction produces an electromagnetic shower electron. Whereas $\nu_{\mu}$ charged current interactions produce a muon as a narrow track along its trajectory rather than shower (Fig.\ref{fig:evd}). The more difficult is neutral current interactions with a single $\pi^{0}$. The $\pi^{0}$ decays to two photons and these photons produces electromagnetic showers that can be difficult to distinguish from electrons. Photons travel some distance (the photon conversion distance is $\sim$ 38 cm (6 planes)) before converting into an $e^{-}/e^{+}$ pair which produce scintillation light. Therefore we have developed different kinds of reconstruction tools in the NOvA experiment for different purposes. The full chain of reconstruction is outlined in the following sections. The NOvA detectors collect raw data from the read out, referred to as cellhits (activity on a particular cell), it saves information about plane, cell, time and charge information about the hits. Spatial and temporal correlation between the hits are clustered together into different groups called ``slices'' which is the foundation for all later reconstruction stages explained in section \ref{sec:slicing}. Next, a modified Hough transform is applied to identify prominent straight-line features in a slice that serve as seeds (Section \ref{sec:multihough}). Then, the Hough lines are used to reconstruct a global 3D neutrino interaction vertex using an Elastic Arms algorithm (Section \ref{sec:vertexing}). The vertex is then used as a seed to a ``fuzzy k-mean'' algorithm that produces prongs (a collection of cell hits with a start point and direction) which contain the activity of particles in the event (Section \ref{sec:fuzzyk}). Using the slicing and fuzzy k-mean algorithm, the Break Point Fitter (BPF) algorithm makes reconstructed 3D tracks under each of the three particle assumptions (muon, proton, and pion) for each Fuzzy-K 3D prong (Section \ref{sec:BPF}). Another tracking algorithm, based on a Kalman filter generates reconstructed tracks from individual slices and the goal of the tracking is to trace the trajectory of individual particles that deposit energy in the detector. This is especially useful in identifying particles that do not create large electromagnetic or hadronic showers, such as muons (Section \ref{sec:kalmanT}). Identification of neutrino interactions based on their topology without the need for detailed reconstruction is done by means of a Convolutional Visual Network (CVN) (Section \ref{sec:evtcvn}). Similarly CVN is used for particle identification (Section \ref{sec:partcvn}).

\section{Isolating Neutrino Interactions \label{sec:slicing}}
NOvA collects data in the form of packet of hits in 550 microsecond  readout windows for the entire detector. However, physics interactions take place within a shorter time period. In the FD which is located on the surface, the primary concern is separating 50-70 cosmic rays in a 550 microsecond readout windows. where as ND is expected to separate $\sim$ 5 neutrino interactions in each 10 microsecond neutrino beam window. The beam spill (the length of time within which we expect neutrino interactions from the beam to occur) is only within the window of 10 microseconds and the time outside the beam spill allows us to do background determination. All those collected hits can be divided into two groups of hits: signal and noise. To separate accurately the signal hits from the noise hits and further separates the signal hits into clusters of hits that originate from different sources, we use an expanding, density-based clustering algorithm (DBSCAN) described in \cite{dbscan} which makes use of space and time information. In the DBSCAN algorithm, there are two types points: core and border. The minimum number of neighbors (pairs whose distance is below a threshold) is defined as core points whereas border points have less than the minimum number of neighbors, but are allowed to be included in the group of hits (clusters) if and only if they are the neighbor of a core point. The algorithm makes clusters by expanding the cluster around the core points. Points that do not belong to these categories are treated as noise. This algorithm takes as input a pair of cell hits and calculates a distance metric in space and time defined as 
\begin{equation}
\epsilon = \left(\frac{|\Delta T| -| \Delta \vec{r}|/c}{T_{\text res}}\right)^{2} + \left(\frac{\Delta Z}{D_{\text pen}}\right)^{2}  + \left(\frac{\Delta X \text {or} Y}{D_{\text pen}}\right)^{2} 
\label{eq:xsec}
\end{equation}
where $T_{\text res}$ is the timing resolution for the quadratic sum of two  hits (time resolution for FD is $\sim$ 10 $ns$ and for ND is $\sim$ $5$ $ns$), $D_{\text pen}$ is a distance
￼￼￼penalty, $\Delta T$ is the time in nanoseconds between hits, $\Delta Z$  and $\Delta X \text{or} Y $ are the distances in 
￼centimeters between hits in each view. For hits in the same view
 $| \Delta \vec{r} | = \sqrt{\Delta Z^{2} + \Delta X\text{or}Y^{2}} $,
while for hits in opposite views $| \Delta \vec{r} |$  = $\Delta Z$.

The slicing algorithm, called ``Slicer'', requires that each slice contain a single interaction. The performance of the slicer is determined from two metrics: efficiency and purity. 
\begin{equation}
\text {Efficiency}  =\frac {\text {Energy from interaction deposited in slice }}{\text{Total energy from interaction deposited in detector}}
\label{eq:efficiency}
\end{equation}

\begin{equation}
\text {Purity}  =\frac {\text {Energy from interaction deposited in slice}}{\text{Total energy in slice}}
\label{eq:purity}
\end{equation}

In far detector cosmic simulations, slicing was found to have an efficiency and purity of 99.3\%, whereas for the near detector neutrino simulations slicing had a efficiency of 94.4\% and a purity of 98.5\% \cite{reco}.

\section{Identifying Lines with Multi-Hough Transform \label{sec:multihough}}
After slicing, the next step is to identify lines in each slice using a modified Hough transform algorithm \cite{hough}.
This algorithm takes as input pairs of points characterized as a straight line passing through them and parameterized in polar co-ordinates ($\rho, \theta$) where $\rho$ is the perpendicular distance from the line to the origin and $\theta$ is the angle between $\rho$ and the x-axis. The algorithm make lines and fits in each detector view separately. The line passing through each pair of hit points in the slice creates a Gaussian smear vote
\begin{equation}
\text{vote} = e^{\frac{-(\rho -\rho_{0})^{2}}{2\sigma^{2}_{\rho}}}  e^{\frac{-(\theta -\theta_{0})^{2}}{2\sigma^{2}_{\theta}}}
\label{eq:hough}
\end{equation}

where $\sigma_{\rho} = \frac{3}{\sqrt{12}}$ , $\sigma_{\theta} = \frac{3}{d\sqrt{6}}$ and d is the distance between the two hits.  A Hough map is created by filling the phase space with votes and the peak in the map is identified as the line of interest. To create new lines to the Hough map, we use an iterative method. First, we remove the last peak results of hough space and from the rest of the list we look for a new hough map. This continues until we can not find any more peaks in a hough space above threshold. The metrics used to check the  performance of this algorithm is the dominant Hough lines that pass and form intersections near the primary vertex of the slice.

The performance for the Far Detector is within an average of 6.9 (NC), 4.1 ($\nu_\mu$ CC), and 2.7 ($\nu_{e}$ CC) cm of the vertex whereas the average distance for the secondary hough line is 9.9 (NC), 8.2 ($\nu_\mu$ CC), and 8.8 ($\nu_{e}$ CC) cm \cite{reco}.

\section{Identification of Vertex using Elastic-Arm\label{sec:vertexing}}
After slicing and the Hough algorithm, the next reconstruction stage is to run an elastic arm algorithm on each slice to find the primary neutrino interaction point. The output of the algorithm is a global 3D vertex point by seeding the lines made by the Hough algorithm. The single point in a slice where the prong arms meet is the vertex. An ``elastic arms" (also called as ``deformable templates'')  is a straight line defined by polar angle $\theta_{a}$ and azimuthal angle $\phi_{a}$ such that the location of the arm originated at ($x_{0}, y_{0}, z_{0}$) with a distance s in Cartesian coordinates is 
\beqa
 x(s) &=& x_{0} + s\sin\theta_{a} \cos\phi_{a}, \nn \\
  y(s) &=& y_{0} + s\sin\theta_{a} \sin\phi_{a}, \nn \\
   z(s) &=& z_{0} +  s\cos\theta_{a}.
  \label{eq:vtx1}
\eeqa
To best describe the event topology, the Elastic Arm algorithm \cite{Earm} finds the parameters ($x_{0}, y_{0}, z_{0}, \vec{\theta}, \vec{\phi}$) by minimizing an energy function of the form
\begin{equation}
\text {E} = \sum_{i = 1}^{N} \sum_{a = 1}^{M} V_{ia} M_{ia} + \lambda \sum_{i = 1}^{N} \left( \sum_{a = 1}^{M}  V_{ia} - 1\right)^{2} + \frac{2}{\lambda_{\nu}} \sum_{a = 1}^{M}  D_{a}
\label{eq:vtx2}
\end{equation}
where $M$ and $N$ are the total number of arms and hits in slice, respectively. $M_{ia}$ measures distance between cell hit $i$ and arm $a$. This is computed as the perpendicular distance from the hit $i$ to the projection of the arm in the detectors two 2D views, given by 

\begin{equation}
\text M_{ia} =  \left(\frac{d^{\text{perp}}_{ia}}{\sigma_{i}} \right)^{2} 
\label{eq:vtx3}
\end{equation}

 $\sigma_{i}$ is a normalized factor (half of the cell depth/$\sqrt{12}$ = 0.9 cm), $V_{ia}$ is the likelihood that hit i associated with arm a is assumed proportional to $e^{(-\beta M_{ia})}$ and the noise is assumed to be a constant factor $e^{(-\beta \lambda)}$
\begin{equation}
\text V_{ia} =  \frac{e^{(-\beta M_{ia})}}{e^{(-\beta \lambda)} + \sum_{b = 1}^{M} e^{(-\beta M_{ia})}} 
 \label{eq:vtx4}
\end{equation}

$\beta$ is range of influence of each arm, $D_{a}$ is a measure of the distance between the vertex and the first hit on arm a, and $\lambda$ and $\lambda_\nu$  control the penalty terms. The first term in equation 6 measures the goodness of fit between the hits and the arms and it minimizes when arm passes through the hits. The second term is a penalty for hits not associated with any arm. The third is a penalty term for arms whose first hit is far from the vertex location. This term is very important while finding a vertex position for NC events where two photons are produced from $\pi^{0}$ after traveling some distance in the NOvA detector. The likelihood for a photon to travel a distance d before converting is proportional to $e^{({−d}{\lambda_{\nu}})}$, where $\lambda_{\nu} = 7/9X_{0}$ (30 cm), leads to a penalty term.
\begin{equation}
\chi^{2} = −2 ln L = 2 \frac{d}{\lambda_{\nu}} 
\label{eq:vtx5}
\end{equation}
For all the verticies, arms are seeded and the directions are scanned and minimize the energy cost function mentioned in equation 6. The minimization heavily depends on the performance of multi-hough algorithm and uses ROOT’s MINUIT class. The fit procedure is initialized with low values of $\beta$ to avoid local minima in the energy cost function, and $\beta$ is gradually tuned up to reach on the final vertex point in the slice.

The vertex resolutions of a events tell us the performances of both Multi-Hough and Elastic Arms algorithms which is 11.6 (about 2 NOvA cells), 10.9, and 28.8 cm for $\nu_\mu$ CC, $\nu_{e}$ CC, and NC events respectively. 
\section{Formation of Prong with Fuzzy k-Means\label{sec:fuzzyk}}
The next step of the reconstruction chain is the formation of prongs (clusters of hits with a start point and direction. We use a possibilistic fuzzy-k means algorithm [\cite{fuzzyk1}, \cite{fuzzyk2}] for assigning a prong membership to each cell hit within the slice. The term ``possibilistic'' means the sum of each hit’s membership across all prongs is not required to be unity, which allows for outlier hits to be treated as noise. This algorithm works very well on separating the noise hits. The ``fuzzines'' allows a hit to belong to more then one prong. Fuzzy-k makes prongs separately in the XZ and YZ views using the cell hits in a slice, so it starts with a 2D view and furthermore it matches between the two views and produces 3D prong. The algorithm considers the vertex from Elastic Arms as the origin of the event in both views of the detector and the cell hits within the slice appear as peaks of deposited energy in a 1-D angular space around that vertex. Uncertainty is assigned as a function of distance from the vertex. The line connecting a cell hit to the vertex forms an angle with respect to z-direction of detector ranging from $-\pi$ to $\pi$. The uncertainty of this angle is modeled after the multiple scattering of 1-2 GeV muons and electrons associated with each cell hit based on its distance from the vertex. To find the prongs in angular space we seeded prong to find minima in densed cell hit, using the density matrix w:
\begin{equation}
w_{k} = \sum^{n}_{i=1} e^{\left(-\frac{\theta_{k} - \theta_{i}}{\sigma_{i}}\right)^{2}}
\label{eq:fuzzy1}
\end{equation}
with \begin{equation}
￼\theta_{k} =−\pi + \frac{k*\pi}{180},  0 \leqslant k < 360
\end{equation}

Associating each cell hit with a prong starts with assuming there is only one prong centered on the densest cell hit region in angular space and then prong centers are added and updated using the iterative method. The distance from each cell hit $j$ to the prong center $i$ is calculated as
\begin{equation}
d_{ij} = \left(\frac{\theta_{i} - \theta_{j}}{\sigma_{i}}\right)^{2}
\label{eq:fuzzy2}
\end{equation}

and the prong membership is assigned with 
\begin{equation}
U_{ij} = e^{-\frac{m\sqrt{a}d_{ij}}{\beta}}
\label{eq:fuzzy3}
\end{equation}
where a is the number of prong centers in the slice. m is a measure of fuzziness of prongs and is set to 2 to allow membership to be shared between prongs. $\beta$ is a normalization factor which represents the expected spread of hits around the prong center. Furthermore, the prong centers are updated with:
\begin{equation}
\theta^{'}_{i} = \theta_{i} + \frac{\sum^{n}_{i=1} \frac{U^{m}_{ij}}{\sigma_{j}} (\theta_{j} - \theta_{i})}{\sum^{n}_{i=1} \frac{U^{m}_{ij}}{\sigma_{j}}}
\label{eq:fuzzy4}
\end{equation}
Prong angles are updated and additional prongs are added until all cellhits have at least a 1\% membership in a prong at the maximum number of prong seeds has been reached. Prongs with significant membership overlaps are merged. Prongs with large spatial gaps that indicate two colinear particles are split.

\begin{figure}[!htbp]
    \centering
        \includegraphics[width=0.9\textwidth]{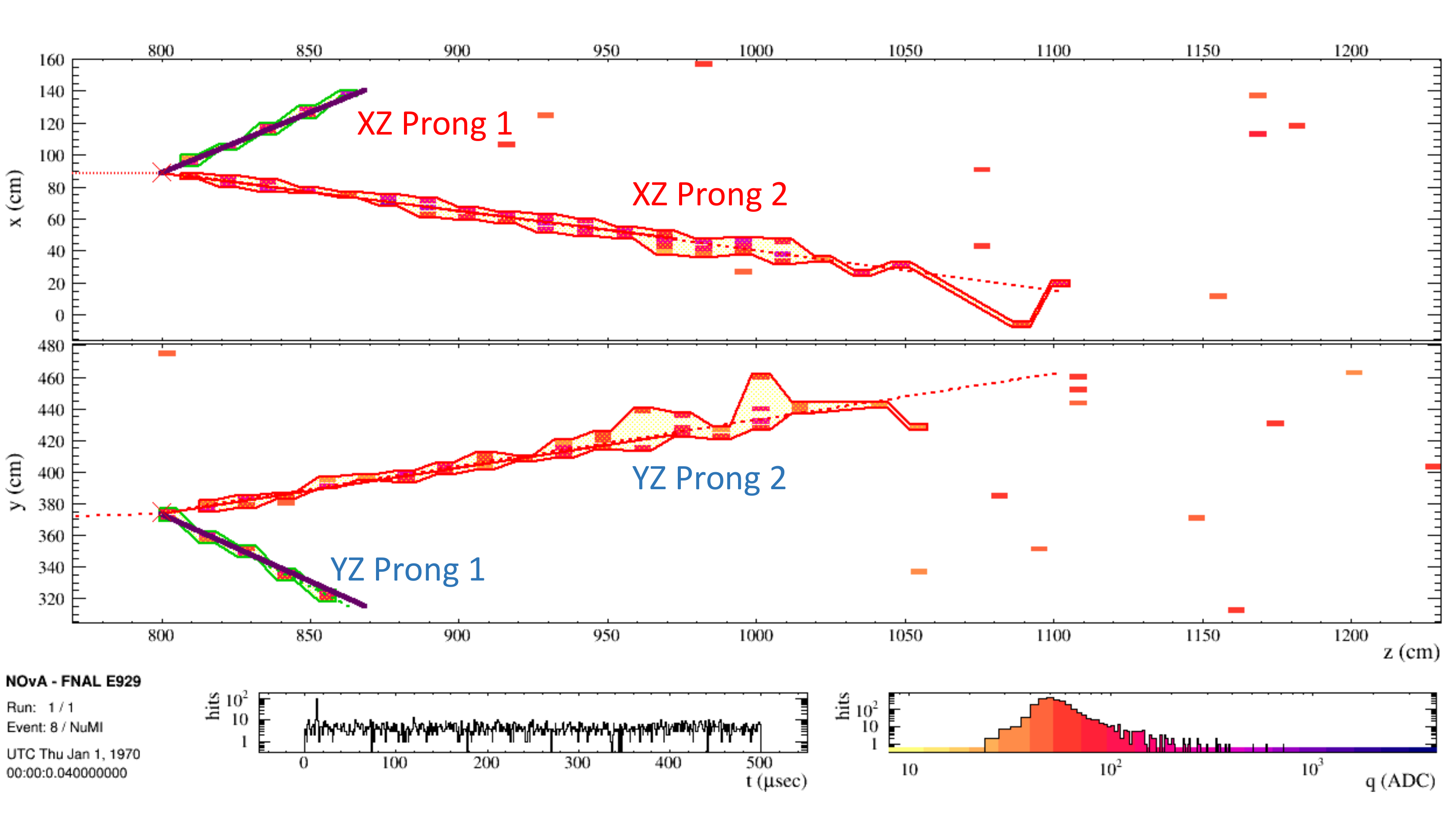}
    \caption{A simulated $\nu_{e}$ CC quasi-elastic interaction in the FD with completed 3D prong reconstruction from the fuzzy-k means algorithm. The reconstructed prong outlined as electron (red) and as proton (green) in each view and the corresponding energy profile histograms used to compute the suitable 3D prong matches is shown in Fig.\ref{fig:fuzzykE} \cite{evan}.}
    \label{fig:fuzzyk}
  \end{figure}

\begin{figure}[!htbp]
    \centering
        \includegraphics[width=0.9\textwidth]{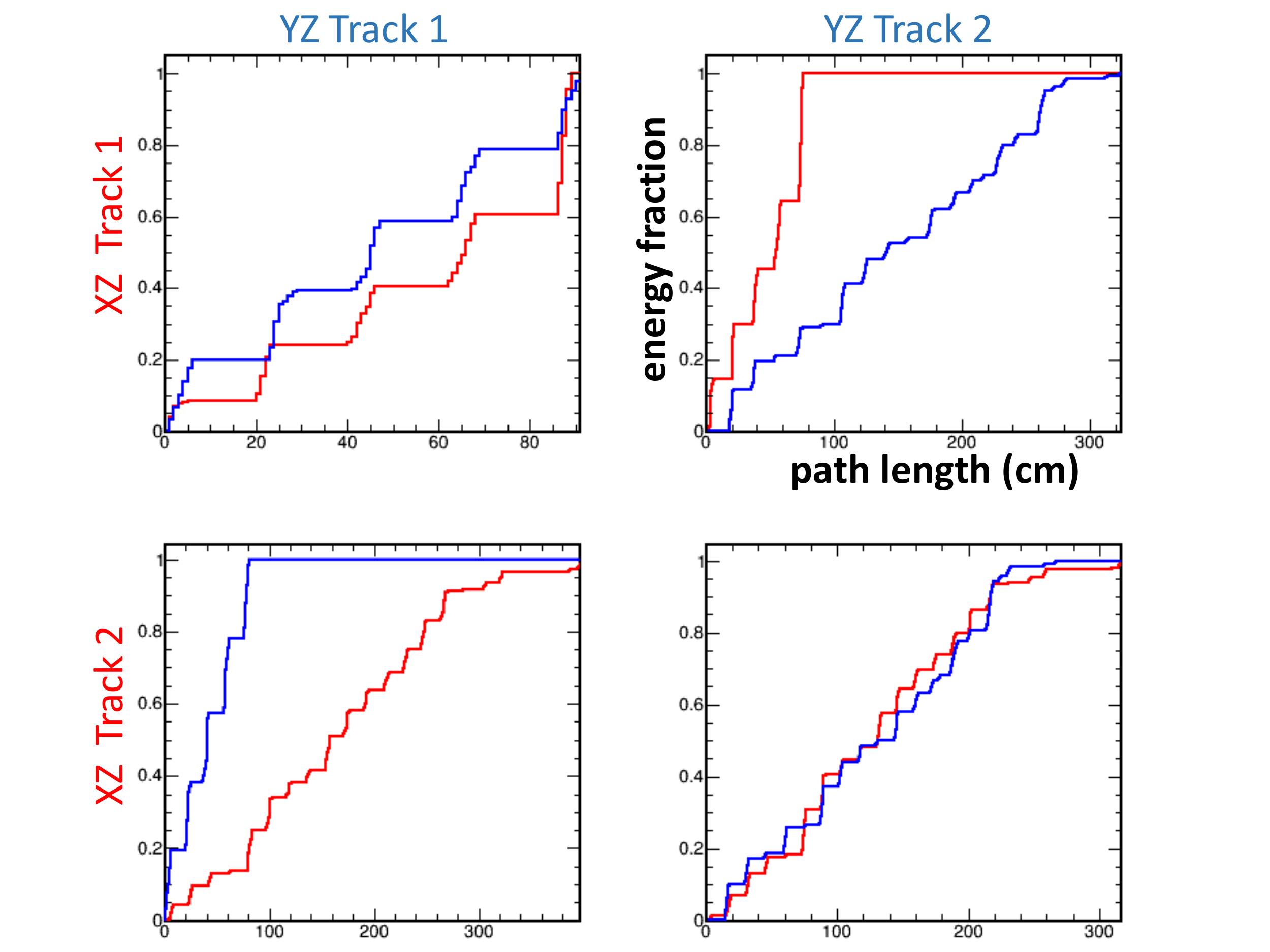}
    \caption{Cumulative energy profile as a function of path length along a prong for perspective 3D match candidates shown in Fig \ref{fig:fuzzyk}. The prongs are in the XZ (vertical planes) view (red curves) and for the YZ (horizonal planes) view (blue curves). The upper-left and lower-right panels show the preferred matches with similar energy profiles that result in the green and red track respectively. The off-diagonal elements illustrate the difference in energy profile shape for the wrong combinations \cite{evan}.}
    \label{fig:fuzzykE}
  \end{figure}

 At the end of the prong formation stage there is a set of 2D prongs for each view of the NOvA detector. The next stage in the process is two match prongs between views in order to form 3D prongs. Matching involves comparing the energy profile of a prong in each view. A Kuiper metric $K = D^{+} + D^{-}$ is used to find  the best match for the prong,  where $D^{+} = \text{max}(E^{XZ}(s) - E^{YZ}(s))$ and $D^{-} =\text{max}(E^{YZ}(s) - E^{XZ}(s))$ are the largest negative or positive distances between the profiles respectively (Fig. \ref{fig:fuzzyk} and \ref{fig:fuzzykE}). 
  
The performance of this algorithm is based on the completeness for hits produced by the primary lepton in charged current (CC) interactions. For $\nu_{e}$ CC events, the average completeness is 88\%: 95\% for quasi-elastic events and 86\% for non-quasi-elastic events. For $\nu_{\mu}$  CC events, these numbers are 93\%, 98\%, and 92\% respectively.
\section{Tracking of Particle with Break Point Fitter\label{sec:BPF}}
Starting with Fuzzy-K 3D prongs and an Elastic Arm vertex, a trajectory of the particle is constructed using multiple coulomb scattering is approximated with a straight line fit and energy loss along the path of each 3D prongs due to deposition of cell hits. The expected energy loss is summed, walking back from the end of the track to the vertex. Moving from the vertex back to the end of the track, scattering planes are inserted based on a multiple scattering model. While walking forward along the trajectory, the expected amount of Coulomb scattering is given by.  
\begin{equation}
\theta_{\text{rms}} =  \frac{13.6 MeV}{\sqrt{3}\beta p} \sqrt{x} (1 + 0.038 \ln (x))
\label{eq:BPF1}
\end{equation}
where $\beta$ is the particle velocity in units of c, p is the particle momentum and x is the distance traveled by a charged particle. When the expected scattering angle becomes more than some tolerance value, we place a scattering plane half a step back. This continues until the end of the track is reached. It is assumed that the kinetic energy of the particle is zero at this point.
\begin{figure}[!htbp]
    \centering
        \includegraphics[width=0.9\textwidth]{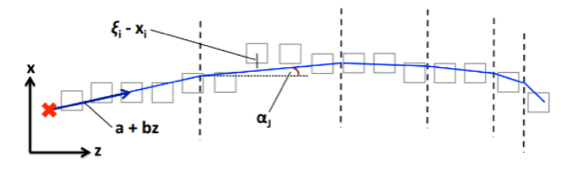}
    \caption{Schematic representation of the track model used by BPF. The track is allowed to scatter through the angles $\alpha_{j}$ at the scattering planes indicated by the dashed vertical lines.}.
    \label{fig:bpf}
  \end{figure}    
Next, the particle’s trajectory is constructed using the Break Point Fitter (BPF) algorithm which first assumes that particles are traversing along the z-axis of an Cartesian coordinate system as shown in Fig.\ref{fig:bpf}. The track location at each position in z has  associated value ($x_{i}, y_{i}$) in x and y direction each with an associated uncertainty $\sigma_{i}$. As the track propagates through the cells, it is allowed to scatter at several no. of locations (m) among the different no. of planes (n) and all these locations of scattering planes are arbitrary. The trajectory of the track at the $i^{\text{th}}$ position and measurement plane $\xi_{i}$ can be expressed as
\begin{equation}
\xi_{i} = a + bz_{i} +\sum^{m}_{j = 1} \alpha_{j}(z_{i} - Z_{j}) \ast\Theta (z_{i} - Z_{j}) 
\label{eq:BPF2}
\end{equation}
where a and b are the intercept and slope of the initial track direction, $\alpha_{j}$ is the scattering
angle (small) at the $j^{\text{th}}$ scattering plane, and $\Theta (z_{i} - Z_{j})$ is the Heaviside function which ensures that only the upstream scattering angles where $Z_{j} < z_{i}$ affect the trajectory at the $i^{\text{th}}$ measurement plane.
To optimize these parameters, we define a $\chi^{2}$ goodness of fit as
\begin{equation}
\chi^{2} = \sum^{n}_{i = 1} \frac{(\xi_{i} - x_{i}) ^{2}}{\sigma_{x_{i}}} + \sum^{N}_{j = 1} \frac{(\beta_{j} - \alpha_{j}) ^{2}}{\sigma_{\alpha_{j}}}
\label{eq:BPF3}
\end{equation}
where $\sigma_{x_{i}}$ is the error on the $i^{\text{th}}$ measurement, $\sigma_{\alpha_{j}}$ is the uncertainty in the scattering angle calculated for each scattering plane j, and $\beta_{j}$ is the expected scattering angle at the $j^{\text{th}}$ scattering plane.

The final results is a reconstructed 3D track under each of the three particle assumptions (muon, proton and pion) for each Fuzzy-K 3D prong.
￼
\section{Kalman Track \label{sec:kalmanT}}
Another useful algorithm which reconstructs tracks from individual slices and is widely used in the NOvA's muon neutrino disappearance analysis is the Kalman algorithm \cite{kalman}. This algorithm  takes input as clusters of hits formed from the Slicer algorithm and forms tracks in the two detector views (XZ and YZ) separately. Each view produces 2D tracks which are later matched to produce a single 3D track. To create 2D tracks, seeding is done where the seed is a segment of the track and formed from  pair of hits that are separated by less than 4 cells. The seed is propagated using a Kalman filter to extend the track and add any additional hits from next cell using the current value of the track position and slope that are consistent with the track. Once a hit is added to the track the position of the track, direction, slope and intercept is updated for the new measurement and the process is continues until no more hits can be added to the track. The propagation process starts from the downstream end of the detector, toward the upstream direction, because in the downstream end the particles emerging from the interaction should be the most separated from each other. Track propagation is continued as long as it did not get any consistent hit and the probability of a gap existing in a track from one hit to the next is less than 0.0001. Once the track propagates to the upstream, propagation is reversed to go downstream to pick up any missing hits from the initial propagation. To find a good track, there is a optimization of track based on maximizing the efficiency of reconstructing long tracks (muons), with rejecting the poor reconstructed track. 

Once all the 2D tracks have been made in each view independently, matching the two views of 2D track is based on the score metic which measures the overlap of 2D track in z-direction in both views. The score metric is defined as:
\begin{eqnarray}
S &=&\frac{\mathrm{Start_{diff}} + \mathrm{Stop_{diff}}}{\text{Length of Overlap in z-direction}} ,\nn \\
 \mathrm{Start_{diff}} &=&  | z_{\text{low of xz track} }- z_{\text{low of yz track}}|, \nn \\
\mathrm{Stop_{diff}} &=&  | z_{\text{high of xz track}} - z_{\text{high of yz track}}|
\label{eq:kalman1}
\end{eqnarray}

Matching starts from lowest value of S and progresses to higher values. The track merging process is performed iteratively until no more 2D tracks can be matched together to form a 3D track. 
 
\section{Classification of Events using CVN \label{sec:evtcvn}}
\begin{figure}[!htbp]
    \centering
        \includegraphics[width=0.45\textwidth]{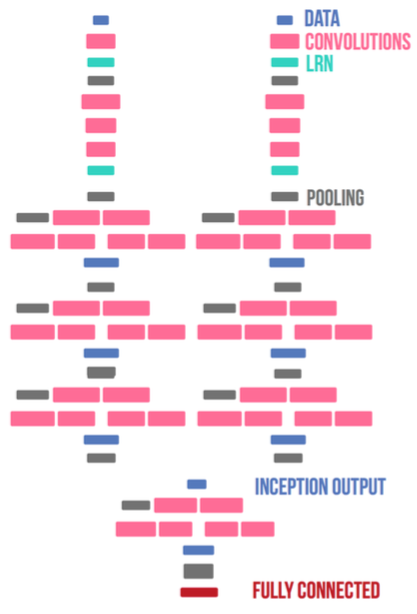}
    \caption{CVN input}
    \label{fig:netlayer}
  \end{figure}

 NOvA's event classification is based on the Deep Convolutional Network in the ``image recognition'' style called Convolutional Visual Network (CVN).  CVN was designed and trained using the  Caffe \cite{caffe} framework which is a collection of libraries and methods to train Convolutional Neural Networks (CNNs). CVN was trained on approximately 4.7 million simulated neutrino interactions and all cosmic ray interactions coming from data taken at the far detector. The network is trained on the two views of calibrated hits in the form of pixel maps (A mapping of data from the detector view into an NxN matrix of values corresponding to pixel contents). The information from each view is then combined in the final layers of the network as shown in Fig .\ref{fig:netlayer}. The convolutional layers apply learned kernels of given dimensions to the image and their output is of the same dimensions as the input image. The pooling layers are functionally identical to convolutional layers but focus on reducing the size of the output. The final stage is the fully connected layer, where the features extracted by the previous layers have been turned into sets of variables and weights which are fed through a multi-layer perception.
  
\begin{figure}[!htbp]
    \centering
       \includegraphics[width=0.45\textwidth]{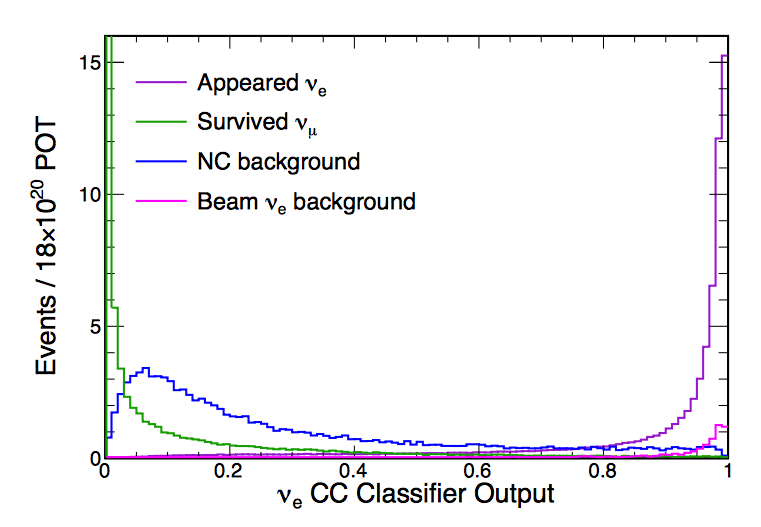}
      \includegraphics[width=0.45\textwidth]{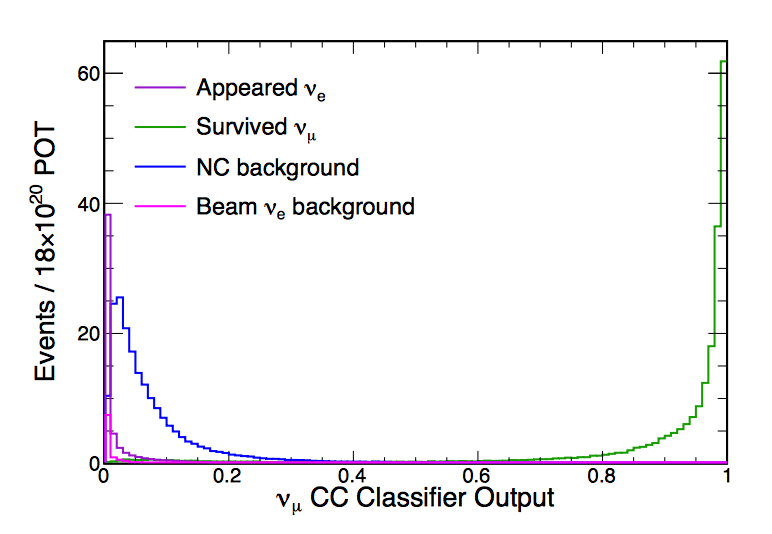}
    \caption{$\nu_{e}$ CC Classifier output (left) and $\nu_\mu$ CC  Classifier output (right) and comparison contains simulated events only for appearing $\nu_{e}$ CC interactions (violet), surviving CC  $\nu_\mu$ (green), NC (blue), and NuMI beam $\nu_{e}$ CC interactions (magenta).}
    \label{fig:classifier}
  \end{figure}

The output of CVN  is a vector of scores  running from 0 to 1 which can be interpreted as a probability of each categories ($\nu_\mu$ CC, $\nu_{e}$ CC, $\nu_\tau$ CC, NC, Cosmic etc.) where 0 is most likely background events and 1 is for signal events. The result of the CVN event classifier for $\nu_{e}$ CC and $\nu_\mu$ CC with the various NuMI beam backgrounds are shown in Fig. \ref{fig:classifier}.
  
Implementation of CVN in the NOvA experiment increased the effective exposure by 30\% compared to traditional event identification  methods in NOvA's $\nu_{e}$ appearance analysis \cite{cvn}.

\section{Classification of Particle using CVN \label{sec:partcvn}}
\begin{figure}[!htbp]
    \centering
       \includegraphics[width=0.7\textwidth]{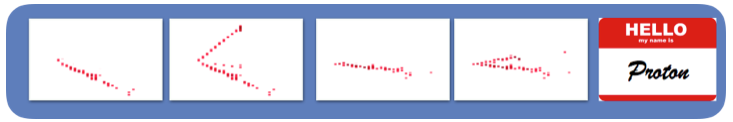}
    \caption{The Schematic diagram shows four views, first two from left are full event in two views and are removed hit from individual prongs.}
    \label{fig:praticle}
  \end{figure}

\begin{figure}[!htbp]
    \centering
      \includegraphics[width=0.7\textwidth]{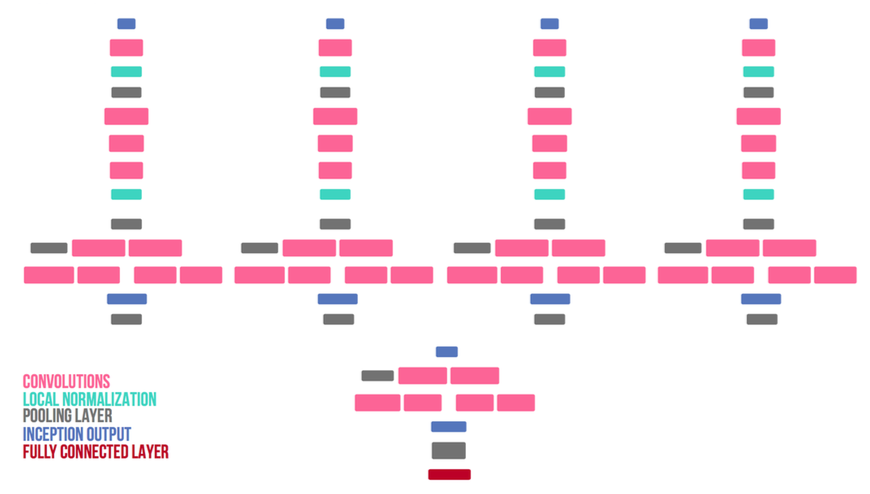}
    \caption{The network is trained on two dimensional views 
   of the event’s calibrated hits and as well as removed hit from individual prongs. At the end the information of each view is then combined in the final layers of the network.}
    \label{fig:prongcvn}
  \end{figure}

Another technique we are developing at NOvA is one which classifies classify individual particles within an event. This effort is ongoing  within NOvA, but I will discuss some initial results. This algorithm is more or less similar to event classification which is completely decoupled from traditional reconstruction chain, but it employs the FuzzyK prong discussed in section \ref{sec:fuzzyk} . These prongs are constructed and matched between views using traditional reconstruction. The CVN network classifies particles based on the contribution of individual particle. The particle classification network uses four views in total, two of the full event and two of the individual prong with the additional hits removed which requires additional layer of network than our usual two layer of event classifier shown in Fig. \ref{fig:praticle} and \ref{fig:prongcvn}. 

\begin{figure}[!htbp]
    \centering
      \includegraphics[width=0.45\textwidth]{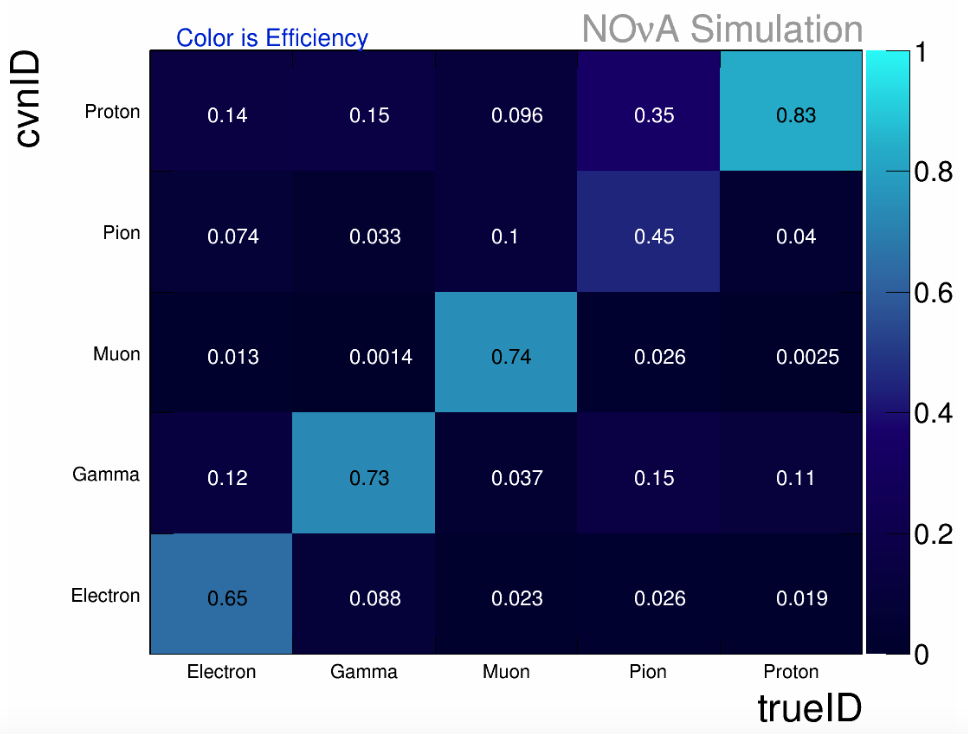}
    \includegraphics[width=0.45\textwidth]{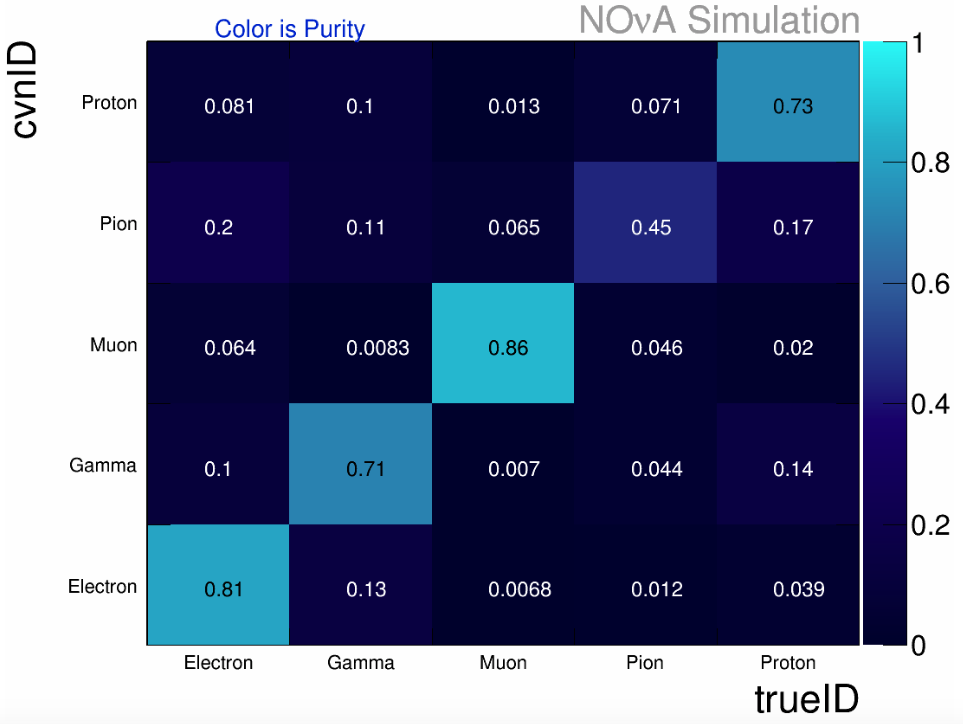}
    \caption{Co-relation matrix between true ID and CVN ID shows for different particles trained using CVN network and the color represent Efficiency (left) and purity (right) and text represent the percentage of co-relation.}
    \label{fig:corelation}
  \end{figure}

 The network was trained on the five different particles (electron, gamma, muon, pion and proton) as shown in Fig \ref{fig:corelation}.  where initial results tells us there is a nice separation of electromagnetic and hadronic activity in the detector.  

\section{Summary}  
The traditional reconstruction methods used by NOvA have proven to be an effective means of identifying particles of interest and reconstructing their kinematics. We have shown that the efficiency of this process can be further enhanced through the use of novel techniques such as machine learning and computer vision. This increased efficiency has resulted in an effective 30\% increase to our neutrino exposure. That number stands to grow as we continue to develop these technologies.

\section{ACKNOWLEDGMENTS}
This work was supported by the US Department of Energy; the US National Science Foundation; the Department of Science and Technology, India; the European Research Council; the MSMT CR, Czech Republic; the RAS, RMES, and RFBR, Russia; CNPq and FAPEG, Brazil; and the State and University of Minnesota. We are grateful for the contributions of the staffs of the University of Minnesota module assembly facility and NOvA FD Laboratory, Argonne National Laboratory, and Fer- milab. Fermilab is operated by Fermi Research Alliance, LLC under Contract No. DE-AC02-07CH11359 with the U.S. Department of Energy, Office of Science, Office of High Energy Physics. 







\end{document}